# Extraordinary room-temperature photoluminescence in WS$_2$ monolayers

*Humberto R. Gutiérrez* [1†], *Nestor Perea-López*[1], *Ana Laura Elías*[1], *Ayse Berkdemir*[1], *Bei Wang* [1], *Ruitao Lv*[1], *Florentino López-Urías* [1‡], *Vincent H. Crespi*[1], *Humberto Terrones*[1,2] *and Mauricio Terrones*[1,3*]

[1] Department of Physics, The Pennsylvania State University, University Park, PA 16802, USA

[2] Departamento de Física, Universidade Federal do Ceará, P. O. Box 6030, Fortaleza, CEP 60455-900, Brazil

[3] Research Center for Exotic Nanocarbons (JST), Shinshu University, Wakasato 4-17-1, Nagano 380-853, Japan

[†]Current address: Department of Physics, University of Louisville, Louisville, KY 40292, USA

[‡]On leave from Advanced Materials Department, IPICYT, Camino a Presa San Jose 2055, Col. Lomas 4a Sección, San Luis Potosí, México

*To whom correspondence should be addressed: mut11@psu.edu*

ABSTRACT: Individual monolayers of metal dichalcogenides are atomically thin two-dimensional crystals with attractive physical properties different from their bulk layered counterpart. Here we describe the direct synthesis of WS2 monolayers with triangular morphologies and strong room-temperature photoluminescence (PL). Bulk WS2 does not present PL due to its indirect band gap nature. The edges of these monolayers exhibit PL signals with extraordinary intensity, around 25 times stronger than the platelets center. The structure and composition of the platelet edges appear to be critical for the PL enhancement effect. Electron diffraction revealed that platelets present zigzag edges, while first-principles calculations indicate that sulfur-rich zigzag WS2 edges possess metallic edge states, which might tailor the optical response reported here. These novel 2D nanoscale light sources could find diverse applications including the fabrication of flexible/transparent/low-energy optoelectronic devices.

MAIN TEXT:

The synthesis and isolation of atomically thin two-dimensional (2D) crystals such as graphene[1] and boron nitride[2, 3] have enabled fascinating advances in 2D metallic (graphene) and insulating (h-BN) systems[4-6]. However, various potential applications require an intermediate behavior, i.e. semiconductors. In this context, monolayers of transition metal dichalcogenides such as $MoS_2$ and $WS_2$, have recently caught the attention of the scientific community as 2D semiconductor crystals with direct gaps in the visible spectrum[7-12]; in bulk form these systems exhibit indirect band gaps. In general, these bulk transition metal disulfides exhibit strong intra-layer sulfur-metal covalent bonds (with metal atoms sandwiched by sulfur layers) and weak interlayer van der Waals stacking. The isolation of single atomic layers from these materials has been an experimental challenge due to their natural tendency to form closed fullerene-like[13-16], nanotube[17, 18], or stacked multilayered[17] geometries. Recently, $MoS_2$ monolayers have been isolated via mechanical exfoliation[7, 8, 19], wet chemical approaches [9, 20, 21], physical vapor deposition[22] and sulfurization of molybdenum films[23]; their electronic and optical properties include carrier mobilities of ~ 200 $cm^2$ $V^{-1}$ $s^{-1}$ and weak room-temperature photoluminescence [7-9, 19, 24, 25]. Regarding $WS_2$ monolayers, they have only been prepared by chemical routes, and their characterization was limited to structural studies[20, 21] and theoretical band structure calculations [10-12].

In this account, we report the synthesis of single- and few-layered 2D triangular micro-platelets of $WS_2$ via the sulfurization of ultrathin $WO_3$ films. These 2D islands displayed extraordinarily high photoluminescence (PL) signal from their edges, at room temperature and in the visible range. A combination of Raman spectroscopy, atomic force microscopy (AFM), scanning and high-resolution transmission electron microscopy (SEM, HRTEM) confirmed that the samples indeed consist of a single S-W-S layer thick, and a giant PL signal occurred from edges exhibiting a zigzag termination. Our first-principles density functional theory (DFT) calculations indicate how the system transitions to a direct band gap material, upon thinning $WS_2$ to an individual layer, and suggest that the sulfur-rich zigzag termination introduces a localized metallic edge state, which could be involved with the strong PL enhancement observed near the edges.

The 2H-$WS_2$ polytype crystalline structure has the hexagonal space group *P6$_3$ /mmc* with lattice parameters of *a*=3.1532 Å, and *c*=12.323 Å [26]. Each $WS_2$ monolayer contains an individual layer of W atoms with a 6-fold coordination symmetry, hexagonally packed between two trigonal atomic layers of S atoms, as depicted in Figure 1a,b. Similar to $MoS_2$, bulk $WS_2$ is an indirect-gap semiconductor, with a gap of 1.3 eV[14, 27]. However, theoretical calculations reveal that with decreasing number of layers, WS$_2$ transitions from indirect (in bulk) to direct (in one individual monolayer) band gap[11], as shown in Figures 1d and 1e (see Supplementary Information for computational details). For bulk WS$_2$, the electronic states involved in the indirect band-gap transition (i.e. valence band maximum at the Γ point and conduction band minimum at the T point) originate from linear combinations of tungsten d-orbitals and sulfur p$_z$-orbitals. These electronic states exhibit a strong interlayer coupling and their energy strongly depends on the number of layers. Consequently, for an individual monolayer, the energy gap between these states (indirect transition) is larger than the direct transition at the K point, thus making the material a direct band-gap semiconductor (Figure 1d and 1e). The conduction and valence electronic states at the K point are mainly due to W d-orbitals and their energies are not very sensitive to the number of layers; the experimental direct band-gap at the K point is *ca.* 2.05 eV [14, 27]. However, small differences between the valence band at the K point for bulk and for a monolayer are noteworthy. In the bulk, there are two well-known direct transitions at the K point due to the splitting of the valence band; experimentally these transitions have been observed by absorption spectroscopy [14, 28] and assigned as excitons A (1.95 eV) [14, 27] and B (2.36 eV) [14, 27] (Figure 1d). However, for a monolayer this splitting is absent (see Figure 1e), and only one direct electronic transition is expected to be observed by optical spectroscopy techniques [28].

One of the most used approaches to produce WS$_2$ nanotubes, fullerene-like structures and films has been the sulfurization of WO$_3$ powders.[29]. We have now modified this process in order to synthesize single- and few-layers WS$_2$ films as well as symmetric triangular clusters of different sizes. First we deposit ultra-thin films (5 – 20 Å) of WO$_3$ on a SiO$_2$/Si substrate by thermal evaporation of WO$_3$ powder under a high vacuum. Subsequently, these films are exposed to a sulfur-rich atmosphere in a quartz tube reactor at ~800°C. The thinnest films of WO$_3$ (5 – 10 Å) create triangular two-dimensional islands that correspond to the initial growth stage of a $WS_2$ monolayer. Figure 2a depicts SEM images of monolayer triangular islands; the spots with darker contrast in the island interiors indicate the nucleation of a second layer. The film thickness was determined by AFM in non-contact mode (for as-grown samples on $SiO_2$, Figure 2b,c), and by HRTEM cross-sectional images (for samples transferred to TEM grids, Figure 2d). The typical height measured by AFM for a $WS_2$ monolayer on $SiO_2$ was ~1 nm, while the spacing between the first and second $WS_2$ monolayers was ~0.6 nm. The ~0.6 nm spacing between $WS_2$ monolayers is in excellent agreement with the theoretical value and that obtained by TEM and X-ray bulk diffraction [26]. The larger AFM-derived spacing between the first monolayer and the substrate is not surprising, since it involves distinct tip-sample and sample-substrate interactions (similar effects have been observed in $MoS_2$ [9] and graphene [30]). Small (in the nm range) and large (micron size) 2D $WS_2$ islands coexist in our synthesized samples; we believe the small monolayer islands correspond to the early stages of growth. These initial 2D islands form by sulfurization of small $WO_3$ clusters and expand laterally as W and S species become available, and diffuse on the substrate surface finding lower energy nucleation sites. Lateral growth could eventually result in the coalescence of neighboring $WS_2$ islands of different crystalline orientation, thus yielding to a large variety of island morphologies, as shown in Figure 1f.

Electron diffraction studies from platelets transferred onto TEM grids reveal single-crystal domains larger than a micron (Figure 3d-f). By transferring the $WS_2$ islands to the TEM grid, we observe skew-stacked and folded $WS_2$ layers (Figure 3a), as indicated in the recorded diffraction patterns shown in Figure 3b. HRTEM images from these regions exhibit the expected Moiré patterns (Figure 3c). High-resolution phase-contrast images from a single-crystal $WS_2$ monolayer (Figure 3f) reveal the honeycomb-like structure of the plain view projection. Electron diffraction of small triangular islands also confirmed that they are single crystals mainly exhibiting zigzag edges (Figure 1h-j). Ideal armchair terminations would alternate S and W atoms, whereas zigzag edges would be terminated by a pure element: either S or W. From a chemical standpoint, sulfur is more likely to be on the edges of metal dichalcogenides clusters, as previously demonstrated by scanning tunneling microscopy (STM) of triangular $MoS_2$ crystals[22]. The presence of intense W and S peaks in energy-dispersive X-ray spectroscopy data (Figure 3g), and a minute amount of O, demonstrate that the $WO_3$ films are efficiently sulfurized and fully converted into $WS_2$.

Room temperature Raman spectra of monolayer $WS_2$ platelets (Figure 2g) show $E_{2g}^1$ and $A_{1g}$ phonon modes at 357 and 419 cm$^{-1}$, respectively. These values are slightly different when compared to those of the bulk $WS_2$ crystal (355 and 421 cm$^{-1}$, respectively). Although the red shift is small (2 cm$^{-1}$), these Raman shifts can be used as a reliable indicator of the presence of a $WS_2$ monolayer. The van der Waals interaction between layers in bulk transition metal dichalcogenides stiffens the lattice, a fact that is consistent with the softening of the $A_{1g}$ mode in the monolayer. However, the stiffening of the $E_{2g}^1$ mode contradicts this trend. This anomalous behavior is not well understood. Lee et al.[19] observed a similar behavior in $MoS_2$ monolayers and proposed that it may be caused by stacking-induced structural changes or long-range interlayer Coulomb interactions[31, 32]. For the 488 nm laser excitation, the intensity of the $A_{1g}$ mode, relative to that of the Si substrate peak, monotonically decreases with decreasing the number of $WS_2$ layers. Figure 4a and 4b show SEM images of two different samples, one containing small (~5 μm size) triangular $WS_2$ islands (Figure 4a) with various second- and few-layers plateaus on the surface (darker contrast spots), and the other one (Figure 4b) having a larger monolayer $WS_2$ island (~15 μm size) with only two small regions of higher plateaus located at the center. The Raman mappings of the intensity ratio $I_{A_{1g}}/I_{Si}$ corresponding to those regions are shown in Figure 4c and 4d. The blue region with values of $I_{A_{1g}}/I_{Si}$ ~ 0.5 corresponds to the monolayer part of the islands; this was also confirmed from the frequency shifts of $E_{2g}^1$ and $A_{1g}$ phonon modes. The color pattern in the Raman mappings is very convenient in order to correlate other properties such as PL response with the number of layers and position in the 2D island as we discuss below.

The PL mappings shown in Figure 4e and 4f, correspond to the spatial distribution of the PL maximum intensity when the excitation laser is scanned over the sample surface. The correlation of PL mappings, Raman mappings and SEM images demonstrates that the extraordinary PL signal arises mainly from monolayers of $WS_2$; regions with more than one layer (e.g. P1) only exhibit very weak or the absence of PL. This is to be expected, since few-layer and bulk $WS_2$ exhibit an indirect band gap. Figure 4i shows the PL spectra obtained from five different positions within the large $WS_2$ islands shown in Figure 4b. Since for this excitation wavelength (488nm), the $A_{1g}$ Raman peak intensity decreases with the amount of $WS_2$ material, and both Raman and PL intensity are also equally affected by external factors such as the local electric fields and the laser excitation power [8]; we report the PL intensity normalized to the $A_{1g}$ Raman peak intensity. Other than the Raman peaks, a single sharp PL peak (FWHM ~ 42 to 68 meV) is the only feature observed from 490 to 900 nm. The presence of this single PL peak is in agreement with the theoretical prediction that only one direct electronic transition at the K point should be observed for a $WS_2$ monolayer (Figure 1a). Direct electronic transitions in $WS_2$ originate from excitonic radiative relaxation[14, 28, 29], and for this reason the PL signal always appears at energies slightly lower than the 2.05 eV direct band gap of $WS_2$. The position of the PL maximum varies for different locations on the island, between 1.99 and 1.94 eV. Since the Raman frequencies – which are strain-sensitive – are homogeneous across the entire monolayer region of the sample, we rule out the occurrence of inhomogeneous strain as a possible cause for the inhomogeneous exciton energy. Alternatively, lattice defects such as impurities or vacancies could localize excitons and thereby create inhomogeneities in the exciton binding energies.

The most striking result is the position dependence of the PL intensity even within the monolayer regions. In Figure 4i it can be noticed that the $I_{PL}/I_{A_{1g}}$ ratio can vary from 36, at the interior of a monolayer platelet (P2 in Figure 4), to 880 near the edge (P5). Splendiani *et al.*[8] reported only a three-fold enhancement in mechanically exfoliated $MoS_2$ monolayers. For comparison, we have also synthesized monolayer MoS2 triangular islands and these exhibit weak PL signals, similar as those reported by Splendiani *et al.*[8] For small triangular $WS_2$ islands, shown in Figure 4e, the maximum PL intensity is also obtained close to the borders and the vertices of the islands. The origins of the hot spots in the PL signal are still unknown, however, the spatial distribution of the PL peak energy-shift calculated from the PL mappings (Figure 4g and 4h), reveals that the regions with larger red shift are preferentially located close to edges and hot spots. In bulk 3D semiconductors (e.g. Si and GaAs) the interaction of shallow impurities with free excitons forms bounded excitons with higher binding energies. In our case this mechanisms is clearly related to the presence of some specific type of edges. When a large island was scratched mechanically along two parallel lines as shown in Figure 5a, the edges created mechanically do *not* show any significant PL enhancement (Figure 5b). This result suggests that the edge chemistry of as-grown monolayer $WS_2$ islands, is crucial when observing the localized PL enhancement. Dangling bonds or different edge passivation in metal dichalcogenides could produce different spins states that change the edge magnetic properties[33]. Furthermore, recent first-principles calculations have indicated that a high spin density could be localized surrounding metal vacancies in similar types of metal dichalcogenides[11]. The interaction of the free carriers (electrons and holes) with either localized electrical charges (i.e. metallic borders) or localized spin moments, could affect the dynamics of the excitons generated in their vicinity, thus affecting the binding energy as well as enhancing the radiative recombination of electrons and holes. The actual mechanisms leading to the edge-enhanced PL is still to be determined, however, understanding the physical properties of the different kind of edges through first principle calculations and GW quasiparticle corrections can shed light and point to possible directions to solve the problem.

We carried out DFT spin polarized simulations on individual triangular zigzag-edge WS$_2$ clusters S$_{84}$W$_{28}$ and S$_{104}$W$_{36}$ (see Supplementary Information). Two main types of clusters were studied: clusters with sulfur saturated edges and clusters with non-saturated portions at the edges. The sulfur saturation of the metal terminated clusters is a reasonable assumption since the experiment was carried out in a sulfur-rich atmosphere. The interior of the cluster behaves electronically very similar to bulk WS$_2$, while the zigzag edges introduce states at the Fermi level which turn the edges into metallic nanowires (Figure 6a), as previously found in MoS$_2$ clusters [34]. Another feature obtained from our calculations is that for the small WS$_2$ clusters saturated with sulfur, no magnetic features were observed. While, the same clusters with bare W edges (non S-saturation) exhibit small localized spins at the W atoms positioned along the edge (Figure 6b). In order to study the S-saturation effect in extended edges and to make a uniform edge type, we have performed simulations on WS$_2$ nanoribbons. These calculations indicate that the spin polarized configuration of S-saturated zigzag edges in WS$_2$ nanoribbons (Figure 6c), are slightly more stable than the non spin polarized configuration (Figure 6d). Spins at the bare W edge (with no S-saturation) are larger in magnitude. Armchair bare edges in WS$_2$ nanoribbons do not possess states at the Fermi level. Since the synthesized clusters could be several microns in size, the edges should behave similarly to those of $WS_2$ nanoribbons. It is noteworthy that the difference in energy between the ferromagnetic case (the most stable found in our studies), and the non-magnetic configuration is 0.074 eV/unit cell (The unit cell contains 20 atoms). These calculations indicate that the chemistry of the edge is very important because they could show metallic behavior and/or localized spins which might produce a net magnetic moment at the edges. Certainly, GW quasiparticle calculations might add more information on the behavior and nature of the excitonic effects. However, these simulations are computing costly when considering supercells of $WS_2$ sulfur saturated nanoribbons or finite cluster systems [35]. Further experiments are underway in order to determine the actual origin of the enhanced PL, however, the interaction of excitons with the local electric and/or magnetic fields, generated at regions close to the edges, could be one possible explanation for the giant PL observed in our experiments.

In summary, we have reported the first successful synthesis of individual (S-W-S) monolayers of $WS_2$. We also established a Raman signature for individual $WS_2$ monolayers. The observation of room-temperature PL in $WS_2$ monolayer is reported here for the first time. This photoluminescent behavior can be a result of the transition from indirect (bulk) to direct (monolayer) band-gap, which was also predicted in our calculations. Moreover, the PL signal suffers an extraordinary enhancement toward the edges (and corners) of our triangular platelets. This edge-induced PL enhancement has never been reported before and represents a record in two-dimensional metal dichalcogenides. Our observations and theoretical simulations suggest that edges and defects within monolayers of metal dichalcogenides materials could be engineered to tailor their optoelectronics properties. The fact that this material can be synthesized on $SiO_2/Si$ substrates, following a simple and reproducible route, opens up numerous possibilities for real 2D device fabrication. This method was also used in our laboratory to produce thin films of other metal dichalcogenides such as $WSe_2$, $MoSe_2$, $MoS_2$, $NbS_2$ and $NbSe_2$ with diverse electronic properties that are currently under investigation.

FIGURES:

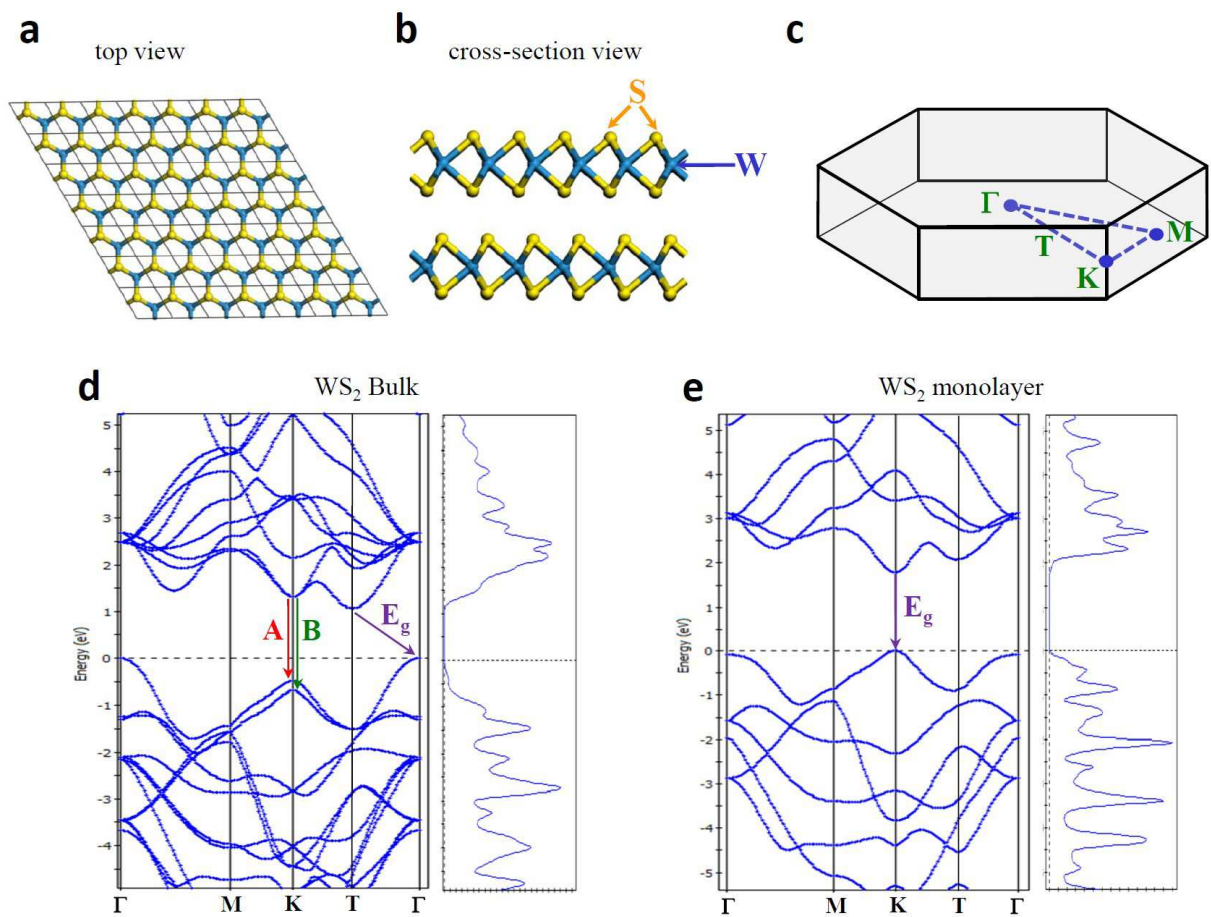

**Figure 1.** (a & b) top and cross-section view of the $WS_2$ atomic structure. (c) Brillouin zone for $WS_2$ monolayer. (d) & (e) Electronic band structure (right) and Total density of states (left) for the $WS_2$ bulk and monolayer, respectively.

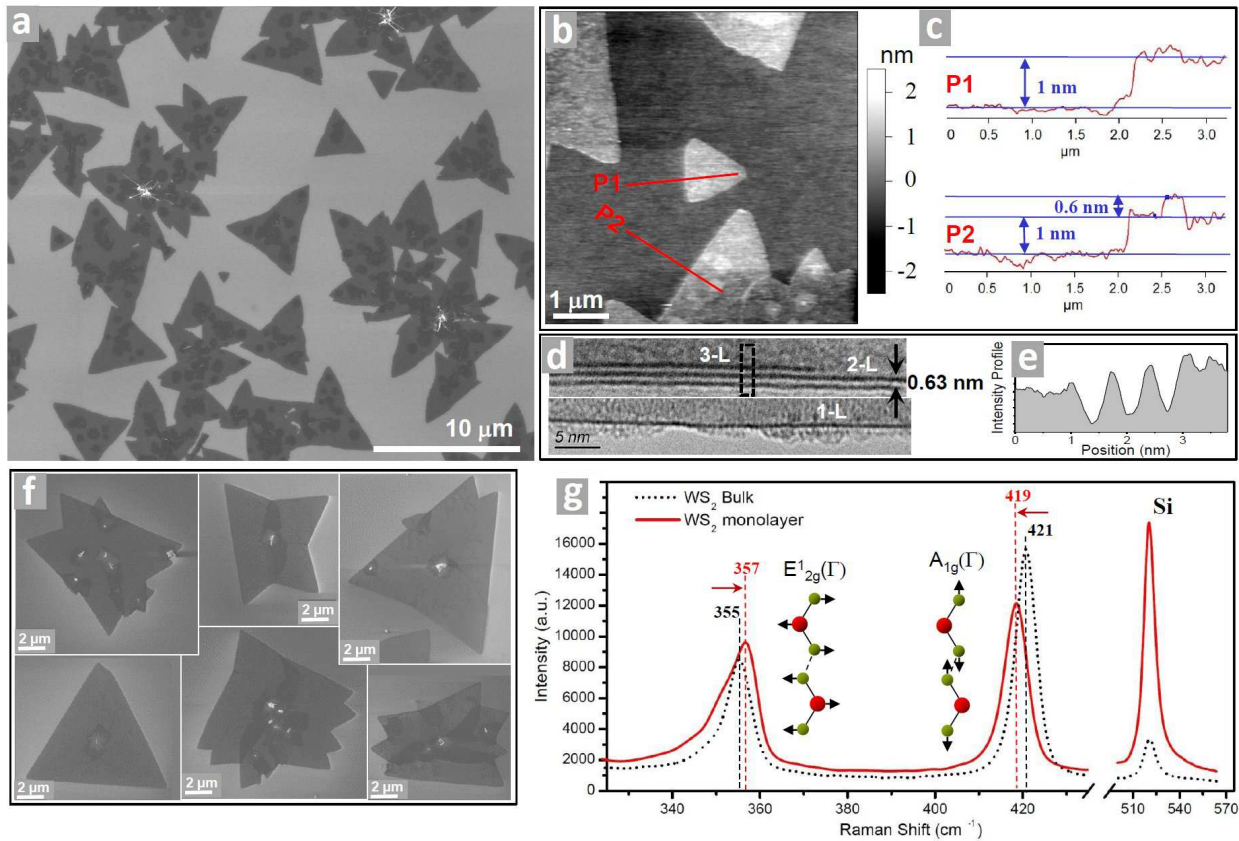

**Figure 2.** (a) SEM image of small triangular 2D clusters of $WS_2$ monolayer, the dark contrast spots in the interior of the islands are second layer clusters in the early stage. (b) & (c) AFM image and height profiles along two different directions (P1 & P2) revealing the positions of one and two $WS_2$ layers. (d) cross-section TEM images of clusters containing 3L, 2L (top) and 1L (bottom), the plot in (e) correspond to intensity profile of the region enclosed by the dasshed box in (d), the spacing between $WS_2$ layers is in agreement with the AFM measurements. (g) is the Raman spectra for $WS_2$ bulk (dotted) and a monolayer (solid red).

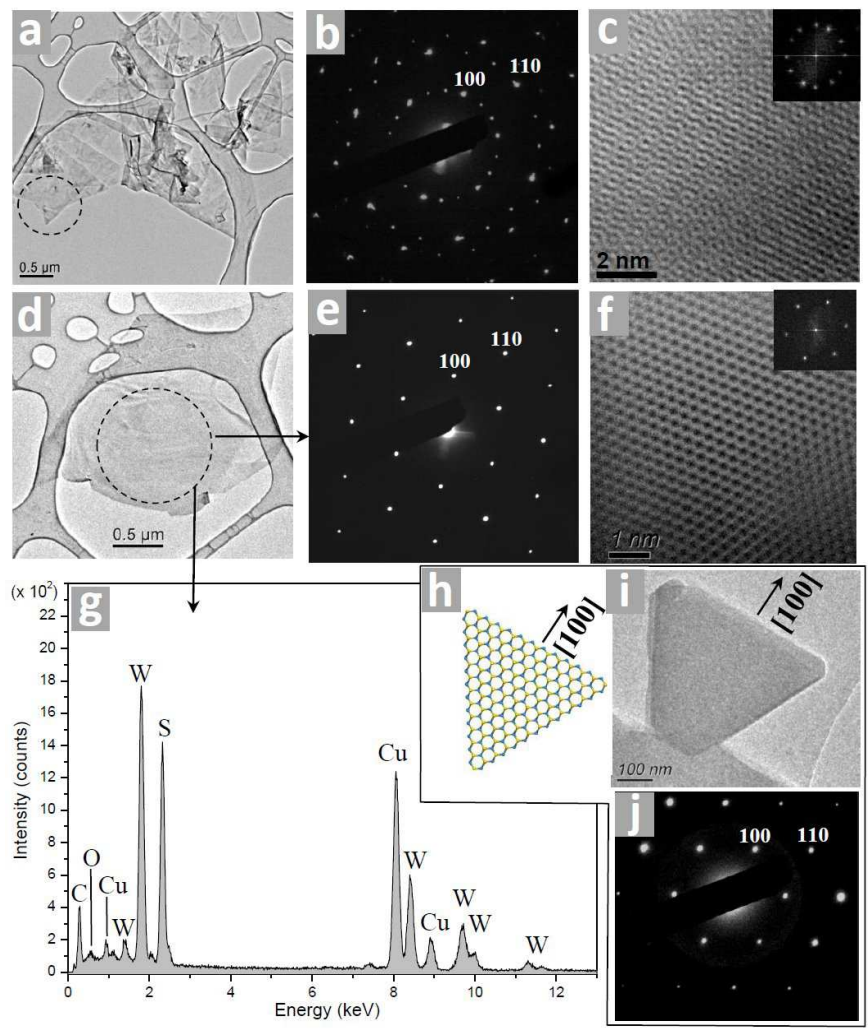

**Figure 3.** (a), (b) and (c) Low magnification TEM, electron diffraction pattern and HRTEM phase-contrast image, respectively, of a $WS_2$ 2D island folded during the transference. (d), (e) and (f) TEM, EDP and HRTEM, respectively, for a single domain 2D $WS_2$ crystal. (g) EDS spectra from the region enclosed in the dashed circle of (d). (h), (i) and (j) atomic model, TEM image and EDP, respectively, of a triangular cluster showing that the edges are perpendicular to the [100] direction (zigzag edges).

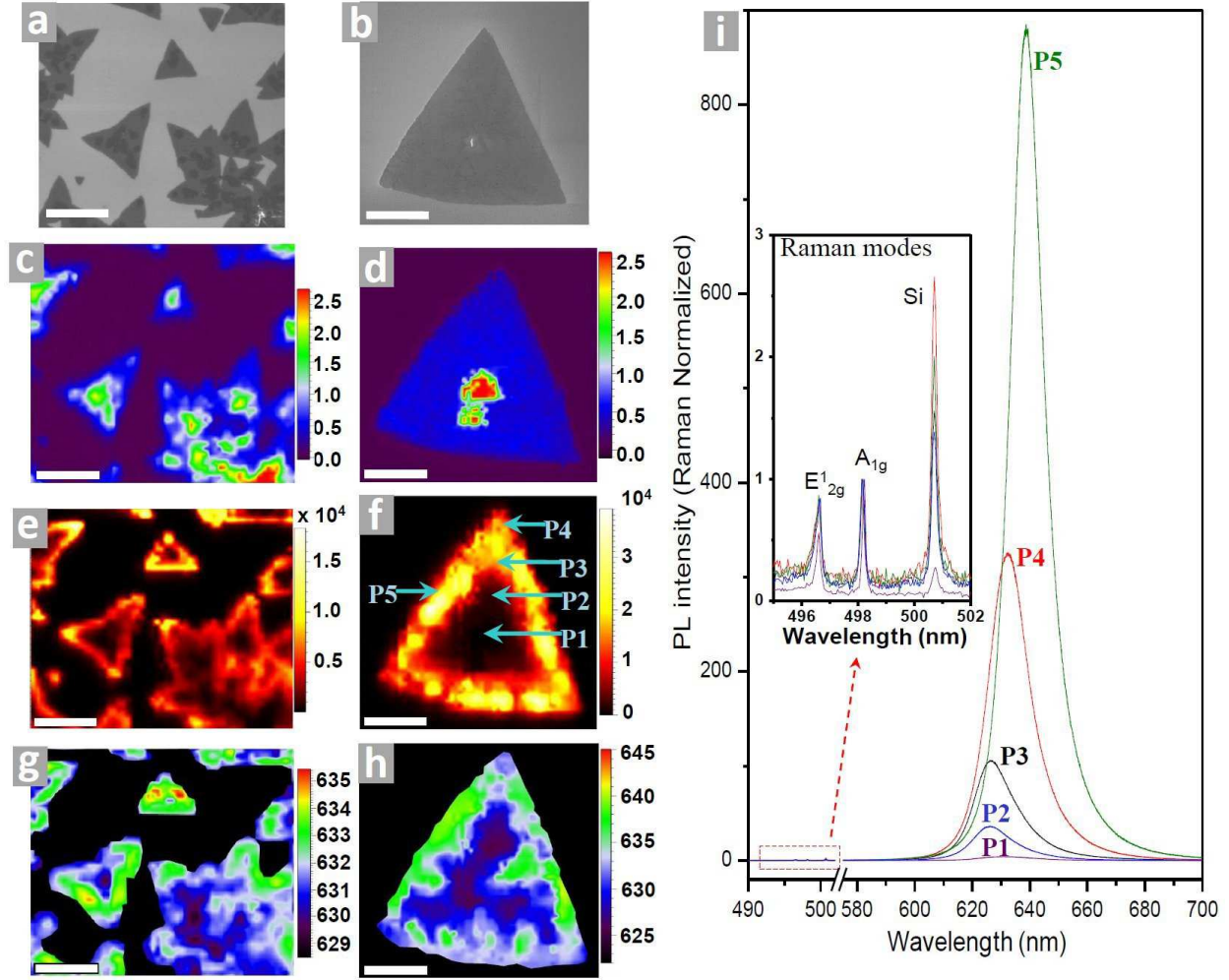

**Figure 4.** (a) & (b) SEM images of small and large islands, respectively. (c) & (d) Raman mappings of the intensity ratio $I_{A_{1g}}/I_{Si}$ corresponding to the islands in (a) and (b), respectively. The region in blue corresponds to a monolayer of WS$_2$. (e) & (f) are the corresponding PL mappings (absolute maximum intensity of the PL peak). (g) & (h) mappings of the PL peak spectral position. The scale bars in (a-h) are 5 μm (i) Room-temperature PL spectra at the different positions in the island indicated by arrows in (f), the PL intensity was normalized to the intensity of the $A_{1g}$ phonon mode ($I_{A_{1g}}$); the inset is a zoom of the Raman peaks.

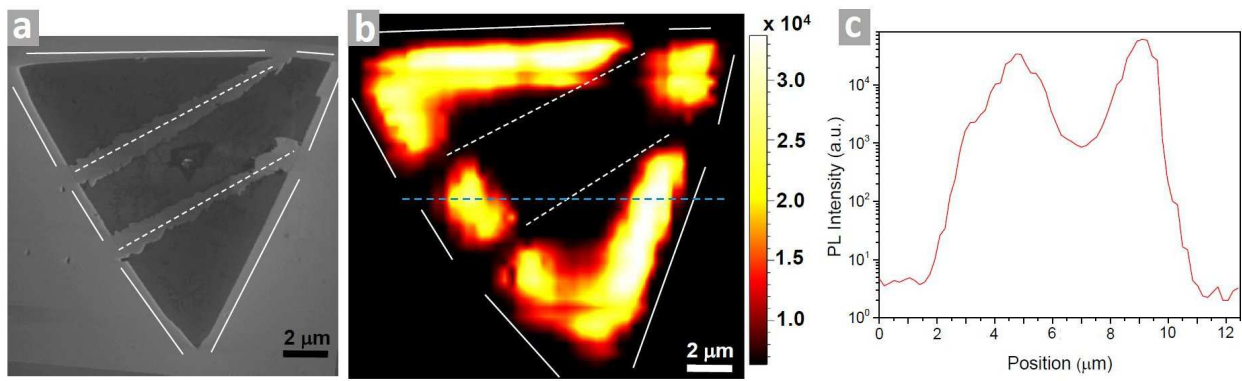

**Figure 5.** (a) SEM image of a mechanically scratched island (mechanical scratches are along the dashed lines) and it corresponding PL map (b). As it can be observed the "new" edges created mechanically do not exhibit PL enhancement. (c) PL intensity profile along the dashed horizontal blue line demonstrating that even at the center of the island there is PL but in less intensity than that obtained close to the edges (notice that the PL intensity scale is logarithmic).

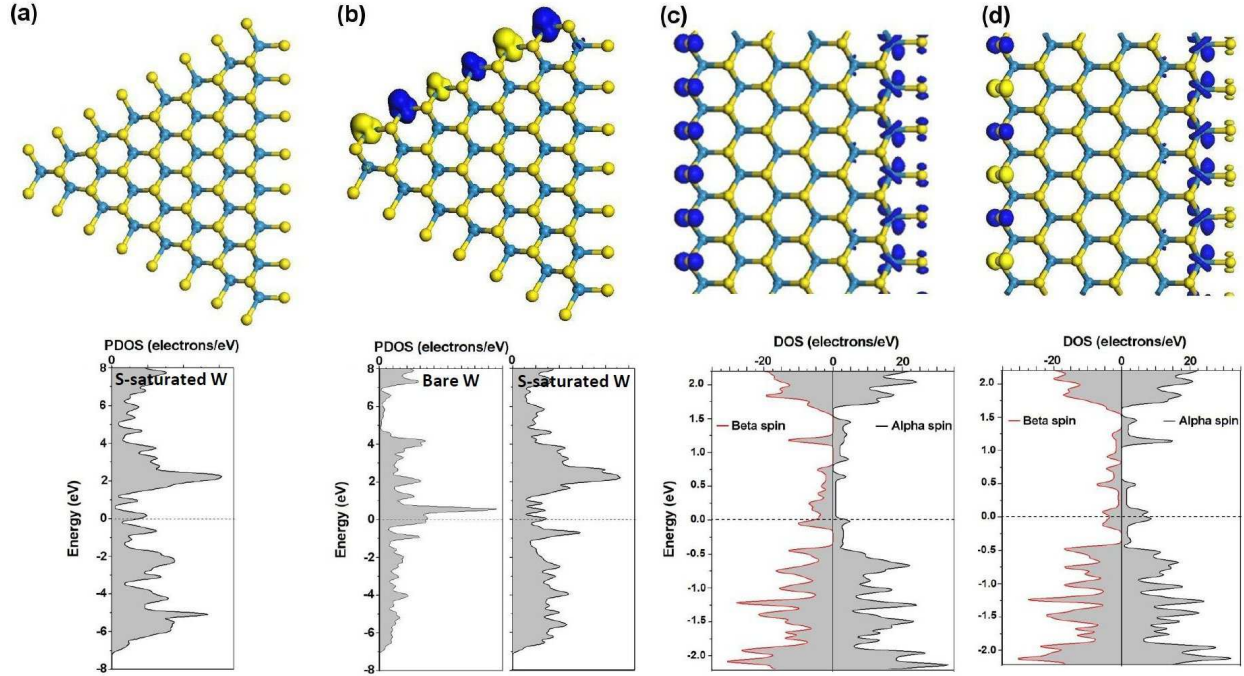

**Figure 6.** (a) (top) Model of the n-8 $W_{36}S_{104}$ saturated with sulfur, and (bottom) PDOS at a W and S atoms at the edges. (b) (top) Model of the n-8 $W_{34}S_{84}$ with one bare edge. Showing the spin density with an iso-surface of $1 \times 10^{-4}$ electrons/atom$^3$, and (bottom) PDOS at the W atoms of the bare edge (left) and a S-saturated W atom (right). (c) (top) Ferromagnetic configuration found for a $WS_2$ 6-zigzag nanoribbon with a magnetic moment of 0.6 $\mu_B$/cell and (bottom) the corresponding spin polarized DOS. (d) (top) Magnetic state found in which there is an anti-ferromagnetic coupling along the sulfur bare edge with a magnetic moment of 0.08 $\mu_B$/cell, and (bottom) the spin polarized corresponding DOS. The energy difference between the states magnetic states c) and d) is 0.02952 eV/cell (the cell in this case is double the unit cell and contains 40 atoms). The ferromagnetic c) state is more favorable.

ASSOCIATED CONTENT:

**Supplementary Information:**

**Materials and Methods:**

**Synthesis.** $WO_3$ thin films (5-20 Å) were thermally deposited on $SiO_2$/Si substrates in a high-vacuum chamber ($10^{-5}$-$10^{-6}$ Torr). Subsequently, the films were transferred into a quartz tube reactor. During the sulfurization of the $WO_3$ films, the samples were kept at 800°C for 30 min under an Argon flow, and S vapors were generated from S powders placed up-stream at a lower temperature region (~250 °C) which was controlled independently.

For the TEM observations, we transferred as grown $WS_2$ islands into gold Quantifoil® TEM grids (from SPI), which have a polymer thin film with 2µm periodic holes. The $WS_2$ islands were released from the original Si/$SiO_2$ wafer by the etching effect of a KOH 15M solution. A first approach [S1] consisted on coating the wafer with a PMMA solution (495k) by spin coating, with a speed of 3000 RPM, during 30 seconds. The polymer was then allowed to cure overnight at room temperature. The edge of the wafer was marked with a sharp blade in order to expose the Si/$SiO_2$ surface, and the wafer was subsequently immersed in the KOH 15M solution. The PMMA/$WS_2$ film was released by the effect of the caustic solution and was fished out with the TEM grid. The TEM grid was placed on absorbent paper and washed with deionized water thoroughly. Finally, the PMMA was dissolved with acetone droplets. A direct PMMA-free approach was also followed [S2]. A TEM grid was placed on the Si/$SiO_2$ wafer containing the $WS_2$ islands. One drop of IPA was allowed to dry on the TEM grid. After 10 min, the wafer was immersed in the KOH 15M solution. The grid was released and placed on absorber paper and was washed thoroughly with deionized water.

**Characterization.** The $WO_3$ films were characterized by Raman and PL spectroscopies, performed in a Renishaw inVia confocal microscope-based Raman spectrometer using the 488 nm laser line as the excitation wavelength. The 520 cm$^{-1}$ phonon mode from the silicon substrate was used for calibration. High-resolution transmission electron microscopy (HRTEM) was carriedout in a JEOL 2010F with an accelerating voltage of 200 kV, field-emission source, ultra high resolution pole piece (Cs = 0.5 mm), 1.9 Å Scherzer limit, and equipped with an energy dispersive X-ray (EDX) spectrometer. Two different microscopes were used in our experiments: LEO 1530 FESEM and FEI NanoSEM 630 FESEM. Non-contact atomic force microscopy was performed in a MFP-3D-SA made by Asylum Research.

**Theoretical modeling.** Density functional theory (DFT) spin polarized simulations on individual triangular zigzag $WS_2$-like clusters with sizes n=7 ($S_{84}W_{28}$) and n=8, ($S_{104}W_{36}$) were carried out using the DMOL3 code as implemented in Materials Studio. The general gradient approximation (GGA) was used as exchange-correlation potential; and the Perdue Burke Ernzerhof (PBE) functional with an atomic cut off radius of 4.5 Å [S3, S4]. The clusters were optimized to reach a convergence energy tolerance of $2\times10^{-5}$ Ha/atom with a maximum force per atom of 0.004 Ha/Å. Additional calculations in which the clusters were put in a box have been carried out, with distances between clusters of 35 Å of vacuum to reduce interactions among them.

$WS_2$ nanoribbons were simulated using a plane wave code as implemented in CASTEP [S5], considering single and double supercells (2×1×1 to allow reconstruction at the edges) under GGA-PBE with a Monkhorst-Pack k-point grid with 9×1×1 k-points and a plane wave basis cut off of 500 eV; optimizing the geometry until the total energy reaches $2\times10^{-5}$ eV/atom and the maximum force per atom exhibits values less than 0.05 eV/Å.

For the bulk WS$_2$ and single layer calculations, the CASTEP [40] plane wave code was used under GGA-PBE considering a Monkhorst-Pack grid with 9×9×1 k-points and a plane wave basis cut off of 500 eV; optimizing the geometry until the total energy reaches 2x10-5 eV/atom and the maximum force per atom exhibits values less than 0.05 eV/Å.

References for the Supporting Information (*S1-S5*)

**Author Contributions**

M.T. and H.R.G. conceived the project ideas. H.R.G. and N.P.L. worked together to develop the synthesis approach and obtained the WS$_2$ samples. H.R.G. and A.B. performed optical spectroscopy (Raman and PL) and data analysis. HRG performed TEM measurements and analysis. N.P.L. and A.L.E. provided SEM images. A.L.E. and R.L. worked on the films transfers to a different substrate. B.W. performed AFM measurements. H.T. and F.L.U. made first principles theoretical calculations. H.R.G., H.T. and M.T. wrote the manuscript. V.H.C. contributed with valuable ideas, discussions and manuscript revision. All the authors discussed the results and commented the manuscript.

**Funding Sources**

Army Research Office through MURI program on Novel Free-Standing 2D Crystalline Materials focusing on Atomic Layers of Nitrides, Oxides, and Sulfides.

ACKNOWLEDGMENT:

M.T. and V.H.C. acknowledges funding from the Army Research Office through MURI program on Novel Free-Standing 2D Crystalline Materials focusing on Atomic Layers of Nitrides, Oxides, and Sulfides. H.T. acknowledges funding from Programa Professor Visitante do Exterior - PVE as "Bolsista CAPES/BRASIL". Supported in part by the Materials Simulation Center of the Materials Research Institute, the Research Computing and Cyberinfrastructure unit of Information Technology Services and Penn-State Center for Nanoscale Science. MT thanks JST-Japan for funding the Research Center for Exotic NanoCarbons, under the Japanese regional Innovation Strategy Program by the Excellence. M.T. and V.H.C also acknowledges support from the Penn State Center for Nanoscale Science for seed grant on 2-D Layered Materials. The electron microscopy characterization facilities within the Materials Research Institute at the Pennsylvania State University were used for this research.